\documentclass[11pt]{article} 

\usepackage[french,english]{babel}
\usepackage{amsmath}
\usepackage{amsfonts}
\usepackage{amssymb}
\usepackage{graphicx}
\usepackage{times}

\def\figskip{\vskip .2cm plus 1mm minus 2mm}

\title{Quark Lagrangian diagonalization versus\break non-diagonal kinetic terms} 

\author{Q. Duret
\footnote{duret@lpthe.jussieu.fr}
, B. Machet
\footnote{LPTHE tour 24-25, 5\raise 3pt \hbox{\tiny \`eme} \'etage,
          UPMC Univ Paris 06, BP 126, 4 place Jussieu,
          F-75252 Paris Cedex 05 (France),\newline
Unit\'e Mixte de Recherche UMR 7589 (CNRS / UPMC Univ Paris 06)}
     \footnote{machet@lpthe.jussieu.fr}
, M.I. Vysotsky
\footnote{SSC RF ITEP, lab. 180,
Bolshaya Cheremushkinskaya Ul. 25,
117218 Moscow (Russia)}
 \footnote{vysotsky@itep.ru}
} 
\date{October 24th 2008} 

\begin{document} 

\maketitle 
 
\begin{abstract} 
 
Loop corrections induce a dependence on the momentum squared of the 
coefficients of the Standard Model Lagrangian,
making highly non-trivial (or even impossible)
 the diagonalization of its quadratic part.
Fortunately, the introduction of appropriate counterterms solves
this puzzle. 

\end{abstract} 
 
The Standard Model (SM) Lagrangian is determined by its
field content and two requirements: local
$SU(3) \times SU(2)_L \times U(1)$ gauge invariance and
renormalizability. The kinetic terms of quarks are 
taken canonically normalized and diagonal in flavor, while 
their interactions with the Higgs doublet contain two
$3\times 3$ complex matrices of Yukawa couplings, respectively
for up and down quarks. In this way, after the Higgs field
has acquired a vacuum expectation value, quark mass terms
non-diagonal in flavor appear. Since it is much more convenient 
to deal with fields which have definite masses, one performs 
a diagonalization of the latter with the help of
 four unitary $3\times 3$ 
matrices: $U_L$ and $U_R$ acting respectively
on left- and right-handed up quark fields, and $D_L$ and $D_R$ acting 
analogously on down quarks. As a result, one obtains new quark fields 
with diagonal masses, while the charged quark currents are described 
by the Cabibbo-Kobayashi-Maskawa (CKM) unitary matrix which equals
$U_L^\dagger D_L$. The matrices $U_R$ and $D_R$ drop
out from the final Lagrangian. 
 
This well-known result needs further consideration when radiative 
corrections are taken into account. Let us discuss the case of two 
generations (generalization to three and more generations is 
straightforward). The point is that, due to the diagram shown in 
Fig. 1, the following non-diagonal kinetic term is 
generated
\footnote{We use the convention $\gamma_\mu(1+\gamma_5)$ in the charged
currents of the SM.\newline
Of course, diagonal terms are generated as well 
but, for them, the transformation to canonical fields is
simple: it should be performed at $p^2 = m_i^2$, see below.}: 
\begin{equation} 
L_{sd} = f(p^2, m_u^2, m_c^2, m_W^2)\, \bar d\, \hat\partial\,(1+ 
\gamma_5)s \;\; . \label{1} 
\end{equation} 
 
\vbox{
\begin{center}
\includegraphics[height=4truecm,width=8truecm]{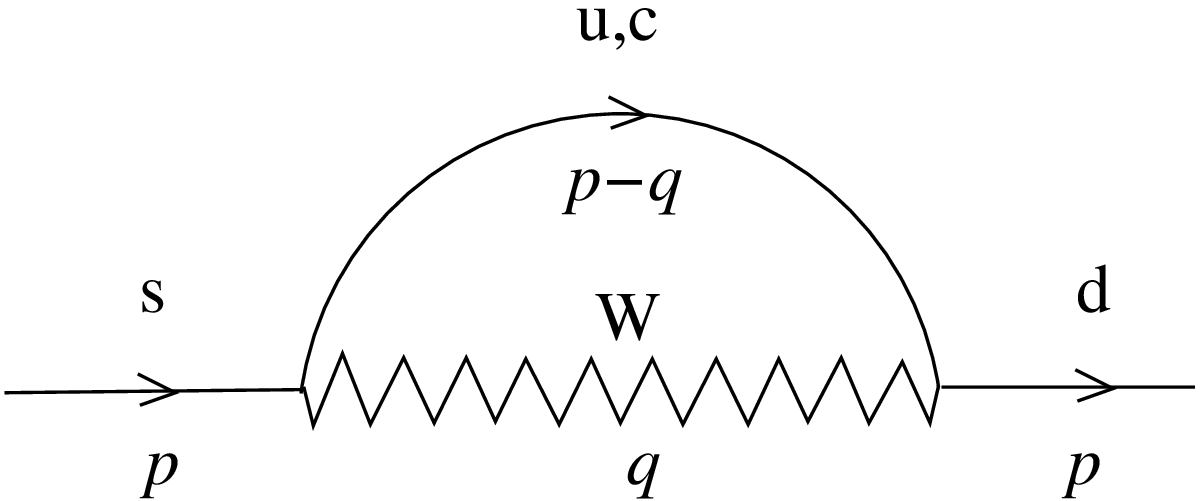}
\figskip
{\em Fig.~1: $s \to d$ transition at 1-loop}
\end{center}
}

As a result, the canonical form of the kinetic terms is lost.
Would we like to diagonalize back the matrix of kinetic terms to
its canonical form (equal to the unit matrix), a  non-unitary 
transformation of quark fields would be needed. 
 
It is not a big trouble if, after such a non-unitary 
transformation, one gets mixed mass terms, since the problem is
then reduced to the one which was already solved:
after  bi-unitarily transforming again the mass 
matrices, one gets quark fields with definite diagonal masses
and non-diagonal charged currents 
described by a unitary CKM matrix. These quark fields are
expressed in terms of the initial (bare) quark fields
through non-unitary matrices. 
 
All this works if the function $f$ does not depend on $p^2$; 
however, it does. 
Thus, one is forced to introduce rotation matrices for
quark fields which depend on $p^2$, and, so, quark mixing angles 
which also depend on the quark virtuality.
For example, different Cabibbo angles then occur
at $s$- and $d$-quark mass-shells. 
 
The obtained result seems far from the standard CKM 
phenomenology, such that a natural question arises: how to implement 
the standard CKM approach to flavor non-diagonal transitions when 
radiative corrections are taken into account? 
 
The solution lies in paper \cite{1}, in which counterterms were
found which, when added to the SM Lagrangian,
cancel the non-diagonal kinetic term (\ref{1}) 
on $s$- and $d$-quark mass-shells (the non-diagonal quark self-energy 
sandwiched between quark propagators has no pole on both quark 
mass-shells). One can easily check that the following counterterms 
do the job: 
\begin{equation} 
L_{sd}^c = -A\, \bar d\, \hat\partial\,(1+\gamma_5) s - B\, \bar 
d\,\hat\partial\,(1-\gamma_5)s - C\,\bar d (1+\gamma_5) s-D\,\bar 
d(1-\gamma_5)s \;\; , \label{2} 
\end{equation} 
where 
\begin{equation} 
A = \frac{m_s^2 f(m_s^2) - m_d^2 f(m_d^2)}{m_s^2 - m_d^2} \; , 
\;\; B = m_s m_d\; \frac{f(m_s^2) - f(m_d^2)}{m_s^2 - m_d^2} \;\; , 
\label{3} 
\end{equation} 
$$ C = -m_s B \; , \;\; D = -m_d B \;\; . $$ 
 
As usual, when counterterms are added to a bare Lagrangian,
its parameters are to be considered as the renormalized ones. 
For the case under concern, diagonalizing its quadratic part
without the counterterms (\ref{2}) and (\ref{3}) yields diagonal
 $d$ and $s$ fields; the same property subsists
at 1-loop because  (\ref{2}) and (\ref{3}) (which, being proportional to the
square of the coupling constant, only play a role  when
calculating loop corrections)
preserve the diagonality of the $d$ and $s$ fields when the term (\ref{1})
is taken into account.
 
In the approach reported above, the transition $s\to d$ does
not occur on $d$- and $s$-quark mass-shells
(it is absent at tree level and its expression (\ref{1}) at 1-loop
 is canceled by the counterterms  (\ref{2}) (\ref{3})).
That is why a question can be raised: 
in the standard approach to Flavor Changing Neutral Currents,
counterterms are not introduced, such that on mass-shell transitions
should be accounted for. For example, in the
calculation of the $s\to d\nu\bar\nu$ decay 
amplitude, one should not only take into account the vertex and box
amplitudes, but also $Z$-boson radiation from external legs:
$s\to s^* Z\to dZ$ and $s\to d^* \to dZ$. By introducing 
the counterterms (\ref{2}) (\ref{3}),  these 
transitions get canceled and one may think that the expressions for the amplitudes 
will change. In reality they do not,  because the covariant 
derivatives should be used in (\ref{2}); in this way, 
the contribution, in the approach without counterterms,
of the non-diagonal  $s\to d$ transition, is reproduced,
in the approach with counterterms, by the
gauge boson terms arising in the latter.

\vskip 3mm
 
{\em Discussions with E.P. Shabalin are greatly acknowledged. We are 
grateful to L.B. Okun for the suggestion to write this short note.}

\end{document}